\RequirePackage{fix-cm}
\documentclass[twocolumn,ngerman,english,5p,preprint]{elsarticle}
\usepackage[T1]{fontenc}
\usepackage[latin9]{inputenc}
\usepackage{array}
\usepackage{rotfloat}
\usepackage{booktabs}
\usepackage{amsmath}
\usepackage{amssymb}
\usepackage{fixltx2e}
\usepackage{graphicx}
\usepackage{nomencl}
\providecommand{\printnomenclature}{\printglossary}
\providecommand{\makenomenclature}{\makeglossary}
\makenomenclature

\makeatletter

\providecommand{\tabularnewline}{\\}

\RequirePackage{ifthen}
\renewcommand{\nomgroup}[1]{%
  \ifthenelse{\equal{#1}{Z}}{\item[\textbf{Superscripts}]}{%
  \ifthenelse{\equal{#1}{X}}{\item[\textbf{Subscripts}]}{%
  \ifthenelse{\equal{#1}{G}}{\item[\textbf{Greek symbols}]}{}}}}

\renewcommand\[{\begin{equation}} 
\renewcommand\]{\end{equation}} 

\usepackage{framed} 
\usepackage{multicol} 

\renewcommand*\nompreamble{\begin{multicols}{2}}
\renewcommand*\nompostamble{\end{multicols}}

\usepackage{siunitx}

\usepackage{xcolor}

\makeatother

\usepackage{babel}
\begin{document}

\begin{frontmatter}{}

\title{An integrated model for the post-solidification shape and grain morphology
of fusion welds}

\tnotetext[t1]{This manuscript has been published in the International Journal of
Heat and Mass Transfer 85 (June 2015), doi:10.1016/j.ijheatmasstransfer.2015.01.144.}

\author[tud,jmbc]{Anton Kidess\corref{cor1}}

\ead{A.Kidess@tudelft.nl}

\author[ucd]{Mingming Tong}

\author[ucd]{Gregory Duggan}

\author[ucd]{David J. Browne}

\author[tud,jmbc]{Sa\v{s}a Kenjere\v{s}}

\author[3me]{Ian Richardson}

\author[tud,jmbc]{Chris R. Kleijn}

\cortext[cor1]{Corresponding author}

\address[tud]{Department of Chemical Engineering, Delft University of Technology,
Julianalaan 136, 2628BL Delft, Netherlands}

\address[ucd]{School of Mechanical and Materials Engineering, University College
Dublin, Belfield, Dublin 4, Ireland}

\address[3me]{Department of Materials Science and Engineering, Mekelweg 2, 2628CD
Delft, Netherlands}

\address[jmbc]{JM Burgers Centre for Fluid Mechanics, Mekelweg 2, 2628CD Delft,
Netherlands}
\begin{abstract}
Through an integrated macroscale/mesoscale computational model, we
investigate the developing shape and grain morphology during the melting
and solidification of a weld. In addition to macroscale surface tension
driven fluid flow and heat transfer, we predict the solidification
progression using a mesoscale model accounting for realistic solidification
kinetics, rather than quasi-equilibrium thermodynamics. The tight
coupling between the macroscale and the mesoscale distinguishes our
results from previously published studies.

The inclusion of Marangoni driven fluid flow and heat transfer, both
during heating and cooling, was found to be crucial for accurately
predicting both weld pool shape and grain morphology. However, if
only the shape of the weld pool is of interest, a thermodynamic quasi-equilibrium
solidification model, neglecting solidification kinetics, was found
to suffice when including fluid flow and heat transfer.

We demonstrate that the addition of a sufficient concentration of
approximately $\SI{1}{\micro\meter}$ diameter TiN grain refining
particles effectively triggers a favorable transition from columnar
dendritic to equiaxed grains, as it allows for the latter to heterogeneously
nucleate in the undercooled melt ahead of the columnar dendritic front.
This transition from columnar to equiaxed growth is achievable for
widely differing weld conditions, and its precise nature is relatively
insensitive to the concentration of particles and to inaccurately
known model parameters.\end{abstract}

\end{frontmatter}{}

\begin{table*}[!t]   \begin{framed}     \printnomenclature   \end{framed} \end{table*}

\section{Introduction}

Welding is a ubiquitous industrial process of great economic and technological
importance \citep{Moos2013Gesamtwirtschaftliche}. Welding processes
involve complex physical phenomena spanning multiple length and time
scales \citep{TongMultiScaleMintweld}. In particular, fusion welding
processes involve phase changes, heat transfer by conduction, convection
and radiation, as well as surprisingly strong fluid flow driven by
Marangoni effects (gradients in surface tension), all of which are
tightly coupled to one another.

Key mechanical properties of alloy welds are related both to \emph{(i)}
their shape and to \emph{(ii)} their grain structure, and thus it
is highly desirable to exert as much control as possible on both during
weld formation. Whereas the shape is mainly determined by macroscopic
phenomena such as heat transfer and Marangoni driven fluid flow during
the melting phase, and can be largely influenced by modifying the
fluid flow through the addition of surface active species \citep{Heiple1982Mechanism,PitschenederDebRoy},
the grain structure is mainly determined by mesoscopic (grain scale)
crystallization phenomena during solidification of the melt, and can
be controlled by the addition of non-melting grain refining particles
\citep{Villafuerte1990Effect,Davies1975Solidification,Villafuerte1995Mechanisms,Bramfitt1970Effect,Park2011Effect}.

Typically, the grain morphology of a post-solidification weld consists
of columnar dendrites, characterized by elongated, tree-like columns
of solid which have grown into the melt. Equiaxed solidification,
where solidification nucleation occurs within the melt away from the
columnar front, is not common, as the thermal gradient ahead of the
solidification front is too large to allow for the necessary undercooling
for equiaxed growth to occur~\citep{Davies1975Solidification,Gaumann1997Why}.
However, the transition of the common columnar solidification mode
to equiaxed solidification is of practical interest. Equiaxed material
is less prone to the unwanted hot-cracking (tears appearing in the
fusion zone near the end of the solidification process \citep{Campbell2012Defects})
and also impedes the undesired segregation of alloying elements to
the central plane of the weld \citep{Villafuerte1990Effect,Villafuerte1995Mechanisms,David1989Correlation}.

Since quantitative experimental research of macroscopic shape evolution
and mesoscopic structure evolution in welds is tremendously difficult
to conduct \citep{Spittle2006Columnar}, there is a need for efficient,
realistic numerical models that can predict both \citep{Mirihanage2013Computational}. 

The majority of previously published numerical studies (e.g. \citep{PitschenederDebRoy,Ha2005Study,Kanouff1992Unsteady,Lei2001Numerical,Zhang2003Modeling})
on macroscopic phenomena during welding have focused on the evolution
of the weld pool shape and temperature up to the end of the melting
stage, and thus neglect further changes in the weld pool shape during
re-solidification after the heat source has been removed. However,
it has been shown that the shape of the weld pool can still change
significantly during the subsequent re-solidification \citep{Ehlen2003Simulation,Saldi2013Effect}.
A proper study of macroscopic weld formation should therefore include
the solidification stage.

Other previous studies have focused on predicting the post-solidification
mesoscopic grain structure of the weld, which is entirely determined
during the solidification stage. In these studies the microstructure
was either studied by a-posteriori analysis, neglecting undercooling
of the melt and the possibility of heterogeneous nucleation sites
(e.g. \citet{Zhang2003Modeling}), or the solidification progression
was modeled using mesoscopic models, while neglecting the influence
of macroscopic phenomena such as fluid flow in the molten metal on
the weld pool shape \citep{Villafuerte1990Effect,Chen20143D,Duggan2012Combined,Duggan2012Integrated,Koseki2003Numerical,Zhan2008Dendritic}.

However, the macroscale and the mesoscale cannot be separated, as
macroscale phenomena such as fluid flow and heat transfer determine
the evolution of solidification on the mesoscale as well. Therefore,
a proper prediction of weld properties requires a combination of both
types of modeling. In this paper we present such an integrated macroscale/mesoscale
model and we show that it can be used to simultaneously predict the
macroscopic shape and the mesoscopic grain structure of a solidified
conduction-mode laser weld. On the macroscale, we compute the heat
transfer and (thermocapillary driven) fluid flow in the weld. On the
mesoscale, the solidification evolution is determined from actual
interface kinetics, rather than interface equilibrium assumptions.
With this integrated model, we investigate the role of fluid flow
during solidification, and the possible alteration of the solidification
mesostructure in a steel alloy laser weld using grain refining particles.

\section{Mathematical formulation}

\subsection{Governing equations}

\begin{figure}
\hfill{}\includegraphics[bb=0bp 0bp 245bp 180bp]{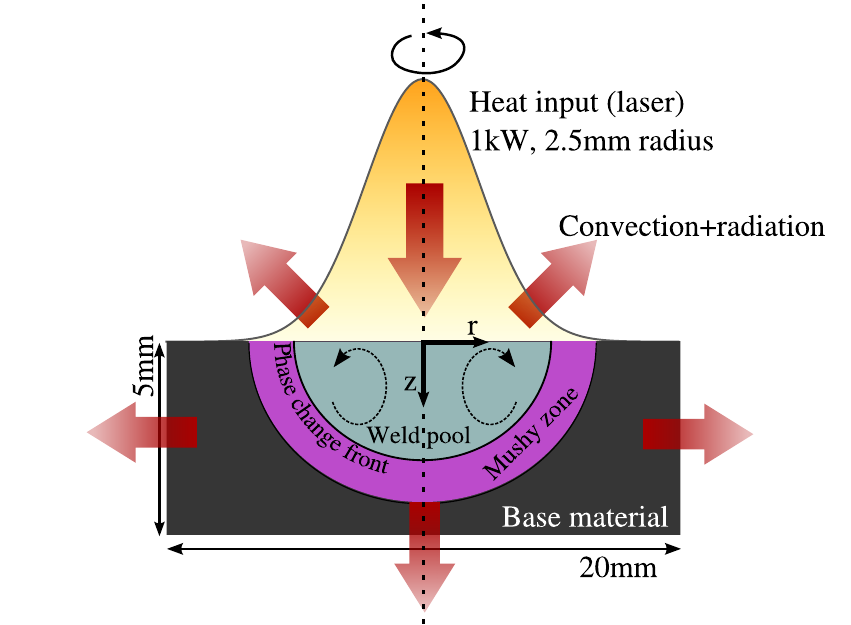}\hfill{}

\protect\caption{Schematic representation of the studied laser welding. The domain
is assumed to be axisymmetric.\label{fig:Laser-welding-schema}}
\end{figure}

A schematic of a typical stationary weld is shown in figure \ref{fig:Laser-welding-schema},
where a non-moving slab of metal is targeted by a fixed high power
laser. The laser irradiation will be absorbed by the target material,
leading to an increase in temperature and eventually a melting phase
change. Heat will be transferred into the bulk of the welded material
by conduction and convection. These phenomena are mathematically modeled
with an energy transport equation with a source term for the latent
heat of the phase change

\begin{equation}
\frac{D}{Dt}(\rho c_{p}T)=\nabla\cdot(\lambda\nabla T)+S_{latent}\label{eq:Energy-equation}
\end{equation}
\nomenclature[aD]{$\frac{D}{Dt}$}{Material derivative}\nomenclature[grho]{$\rho$}{Density}\nomenclature[acp]{$c_p$}{Heat capacity}\nomenclature[aT]{$T$}{Temperature}\nomenclature[glambda]{$\lambda$}{Thermal conductivity}\nomenclature[aS_latent]{$S_{latent}$}{Latent heat source term}

Due to the non-uniform heating of the top surface, large temperature
gradients will develop. These temperature gradients result in gradients
in surface tension, leading to thermocapillary forces along the liquid-gas
interface driving flow in the weld pool. The momentum transport is
described by the Navier-Stokes equations, with a momentum sink that
models the friction in the so-called mushy zone, where the liquid
and solid phase co-exist

\begin{equation}
\frac{D}{Dt}\vec{U}=-\nabla p+\nabla\cdot(\mu\nabla\vec{U})-\vec{F}_{damp}\label{eq:Navier-Stokes}
\end{equation}

\nomenclature[aU]{$\vec{U}$}{Fluid velocity}\nomenclature[ap]{$p$}{Pressure}\nomenclature[gmu]{$\mu$}{Dynamic viscosity}\nomenclature[aFdamp]{$\vec{F}_{damp}$}{Momentum sink term due to solidification}

In the following sections, we will first describe the mesoscale models
used to determine the volume fraction of solid and the developing
mesoscale structure during solidification, followed by a discussion
of the source terms in equations \ref{eq:Energy-equation} and \ref{eq:Navier-Stokes},
and their boundary conditions.

\subsubsection{Columnar dendrite growth}

\begin{figure}
\includegraphics{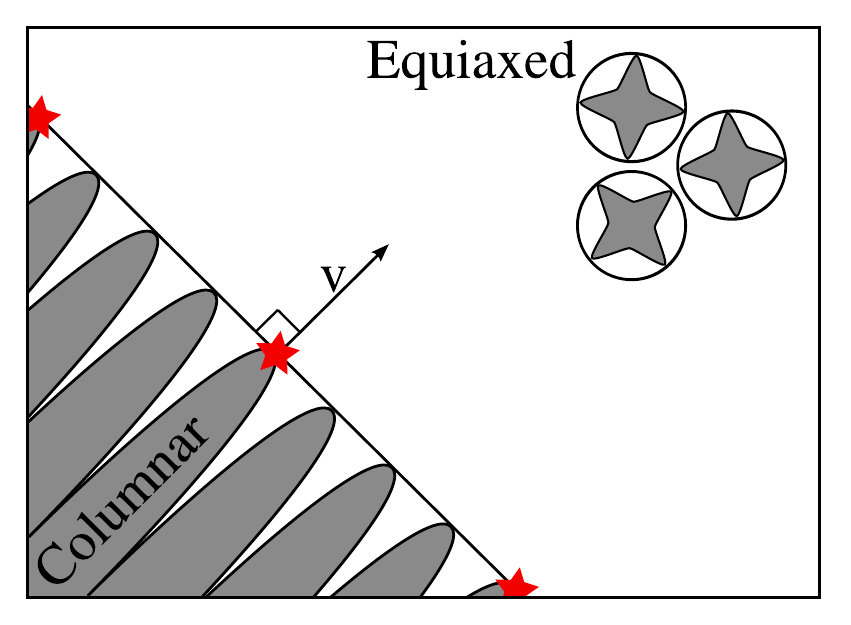}

\protect\caption{Schematic of the mushy zone at the mesoscale, with columnar dendrite
tips, equiaxed dendritic grains and computational markers (stars with
velocity $v$) located at columnar dendrite tips\label{fig:mushy-zone-schematic}}

\end{figure}

At the edge of the weld pool, columnar solidification starts instantaneously
once the melt starts to be undercooled. To compute the growth of columnar
dendrites, we follow the model by \citet{Browne2004Fixed}. Here,
the columnar dendrite front is depicted by a series of massless computational
markers which can be thought of as an envelope that connects all dendrite
tips (see figure~\ref{fig:mushy-zone-schematic}). The markers are
displaced explicitly perpendicular to the local columnar dendrite
front, using an analytically determined local growth velocity: 

\begin{equation}
v=C(\Delta T_{c})^{2}\label{eq:dendrite-radius-increment}
\end{equation}

\begin{equation}
C=\frac{-D_{l}}{8m(1-\alpha)C_{0}\Gamma}\label{eq:BurdenHunt}
\end{equation}
\nomenclature[av]{$v$}{Dendrite tip velocity}\nomenclature[aC]{$C$}{Dendrite kinetics coefficient}\nomenclature[aT_c]{$\Delta T_c$}{Local undercooling}\nomenclature[aDl]{$D_l$}{Diffusion coefficient of solute in the liquid}\nomenclature[am]{$m$}{Slope of the liquidus line}\nomenclature[galpha]{$\alpha$}{Partition coefficient}\nomenclature[aC0]{$C_0$}{Alloy composition}\nomenclature[gGamma]{$\Gamma$}{Gibbs Thomson coefficient}\\
The growth velocity $v$ is dependent on the undercooling $\Delta T_{c}=T_{l}-T$
and a kinetics coefficient $C$, which can be determined using analytical
solutions of dendrite growth. Here, we use the model for $C$ developed
by \citet{Burden1974Cellular} (equation~\ref{eq:BurdenHunt}), which
is based on the hypothesis that structures grow near the optimum condition,
with $D_{l}$, $m$, $\alpha$, $C_{0}$ and $\Gamma$ the diffusion
coefficient of solute in the liquid, the slope of the liquidus line,
the partition coefficient, the alloy composition and the Gibbs Thomson
coefficient, respectively. Alternative growth laws, such as the KGT
model \citep{Kurz1986Theory} based on the marginal stability criterion
\citep{Langer1977Stability}, may easily be incorporated.

The markers are initialized along the liquidus isotherm once the heat
source is extinguished and solidification begins. If a given marker
temperature exceeds the liquidus temperature due to remelting, it
is reset to the closest liquidus isotherm position \citep{Duggan2012Combined}.
Based on the location of markers, a volume fraction of mush $\phi_{col}$
can be determined in a given finite control volume.\nomenclature[gphi]{$\phi_{col}$}{Volume fraction of columnar dendrites} 

\subsubsection{Equiaxed nucleation and growth}

If non-melting grain refining particles are present within the melt,
there is a possibility for nucleation of equiaxed grains within the
weld pool away from the columnar solidification front. Nucleants are
present because they have been purposely added in the form of grain-refining
particles to the weld pool, although in reality they may possibly
also result from fragmentation of columnar dendrite arms. Recent work
\citep{Mirihanage2014Retrieval} has shown that dendrites bend / bow
in the presence of fluid flow. Whether this leads to fragmentation
has not been experimentally confirmed and therefore our model takes
into account added particles only.

For the modeling of the nucleation and growth of equiaxed grains,
we follow a volume averaging approach proposed by \citet{Mirihanage2009Combined}.
The minimum diameter of particles that can act as nucleation sites
for a given undercooling $\Delta T_{c}$ is given by

\begin{equation}
d_{min}=\frac{4\gamma_{sl}}{\Delta S_{v}\Delta T_{c}}\label{eq:Equi-nucleation-diameter}
\end{equation}
\\
where $\gamma_{sl}$ and $\Delta S_{v}$ are the solid-liquid interfacial
tension and the fusion entropy, respectively.

\nomenclature[ad]{$d_{min}$}{Minimum diameter of nucleating particles}\nomenclature[aSv]{$\Delta S_v$}{Fusion entropy}\nomenclature[ggammasl]{$\gamma_{sl}$}{Solid-liquid interfacial tension}

If the grain refiner particles were equisized, we could assume instantaneous
nucleation once the necessary undercooling given by equation \ref{eq:Equi-nucleation-diameter}
is reached. More realistically, however, we assume a log-normal distribution
of grain refiner particle sizes, with a median $\mu_{1/2}$ and a
unitless standard deviation $\sigma_{d}$. The integral of the probability
density function of all particles up to a size $d_{min}$ gives the
cumulative density function

\begin{equation}
F(d_{min})=\frac{1}{2}\text{erfc}(-\frac{\log(d_{min}/\mu_{1/2})}{\sqrt{2}\sigma_{d}})\label{eq:Cumulative-density-function}
\end{equation}

\nomenclature[aFd]{$F(d)$}{Cumulative distribution function for equiaxed nucleation}\nomenclature[gsigmad]{$\sigma_d$}{Standard deviation for the log-normal distribution of inoculant diameters}\nomenclature[gmu12]{$\mu_{1/2}$}{Median inoculant diameter}

The number $N$ of nucleation sites (particles larger than $d_{min}$)
in a control volume $\Delta V$ is then given by 

\[
N=\rho_{seeds}\Delta V\left(1-F(d_{min})\right)
\]

or, in derivative form as

\[
\frac{\partial N}{\partial t}=-\rho_{seeds}\Delta V\frac{\partial F(d_{min})}{\partial t}
\]

\nomenclature[aN]{$N$}{Count of nucleated particles}\nomenclature[grhoseeds]{$\rho_{seeds}$}{Inoculant particle number density}

where $\rho_{seeds}$ is the local number density of grain refiner
seeds (seeds per unit volume). The volume of equiaxed grains $V_{equi}$
within a control volume $\Delta V$ is given by the number of nucleated
grains within $\Delta V$, and the volume of each of those grains.
The change with time in $V_{equi}$ can then be written as

\[
\frac{\partial V_{equi}}{\partial t}=N\frac{\partial V_{ex}}{\partial t}+\frac{\partial N}{\partial t}V_{nuc},
\]

where $\partial V_{ex}/\partial t$ is the growth of volume of existing
grains and $V_{nuc}$ is the volume of newly nucleated grains. The
growth term can be computed as

\[
\frac{\partial V_{ex}}{\partial t}=\frac{4}{3}\pi\frac{d}{dt}R^{3}=4\pi R^{2}v
\]

The dendrite tip velocity $v$ is computed using the same analytical
growth velocity relationship used in the columnar solidification model
(equation \ref{eq:dendrite-radius-increment}). The volume averaged
dendrite envelope radius $R$ at a given time can easily be determined
as\nomenclature[aR]{$R$}{Volume averaged dendrite envelop radius}

\begin{equation}
R(t)=\sqrt[3]{\frac{3}{4\pi}\frac{V_{equi}(t)}{N}}\label{eq:volume_avg_dendrite_radius}
\end{equation}

\nomenclature[aVnew]{$V_{new}$}{Volume of newly nucleated equiaxed dendrites}\nomenclature[aVex]{$V_{ex}$}{Volume of existing equiaxed dendrites}\nomenclature[aVequi]{$V_{equi}$}{Volume of equiaxed dendrites}

From equations~(\ref{eq:Equi-nucleation-diameter}-\ref{eq:volume_avg_dendrite_radius}),
we may now calculate the temporal evolution of the envelope volume
of equiaxed dendrites within the control volume $\Delta V$ around
a location $(\vec{r},\Delta V)$, and from this we may calculate the
local equiaxed grain volume fractions as 

\[
\phi_{equi}^{*}(\vec{r})=\frac{V_{equi}(\vec{r},\Delta V)}{\Delta V(\vec{r})}
\]

After some time the equiaxed grains have grown to an extent at which
they will start feeling the presence of neighboring grains. To model
this so-called grain impingement, we restrict the growth at high volume
fractions using an exponential relationship known as the Avrami equation
\citep{Christian2002Theory}:

\begin{equation}
\phi_{equi}=1-\exp(-\phi_{equi}^{*})\label{eq:grain-impingement}
\end{equation}

\nomenclature[gphi]{$\phi_{equi}$}{Volume fraction of equiaxed dendrites after taking into account grain impingement}

\subsubsection{Grain transport}

Unlike the columnar grains that remain stationary, equiaxed grains
are able to move. Since we assume the grains perfectly follow the
fluid flow, the grain transport can be computed by the solution of
an advection equation for the amount of nucleated grains and the equiaxed
volume fraction: 

\begin{equation}
\frac{dN}{dt}=\nabla\cdot(\vec{U}\, N)\label{eq:grain_transport_equation1}
\end{equation}

\begin{equation}
\frac{d\phi_{equi}}{dt}=\nabla\cdot(\vec{U}\phi_{equi})\label{eq:grain_transport_equation2}
\end{equation}

\subsubsection{Columnar to equiaxed transition}

The columnar front progresses following the growth kinetics given
by equation~\ref{eq:dendrite-radius-increment} until it meets a
coherent equiaxed network. In this case, a columnar to equiaxed transition
(CET) may take place, resulting in a different solidification structure.
To predict this transition, for every time instance we first compute
the growth of equiaxed grains independently of the presence of columnar
grains. The volume fraction of the envelope of equiaxed grains is
clipped in a way that the sum $\phi_{s}=\phi_{equi}+\phi_{col}$ of
equiaxed and columnar dendrites does not exceed 1. Now we compute
the evolution of the columnar dendrite tips using a mechanical blocking
criterion: We assume an equiaxed envelope volume fraction $\phi_{s}$
larger than a certain threshold to be coherent in a way that it mechanically
blocks further advancement of the columnar dendrite tips. Equiaxed
grains that are not bound in a coherent network are integrated into
the columnar dendritic network without blocking its progression. There
is some variance in the published literature on the choice of the
threshold value, ranging from 0.2 \citep{Biscuola2008Mechanical}
to 1 \citep{Mirihanage2010Macroscopic}. Since we neglect solutal
effects, the mechanical blocking is the only mechanism by which a
columnar to equiaxed transition may take place.

\subsubsection{Latent heat release}

The effect of melting and solidification on the heat transfer are
taken into account via the source term $S_{latent}$ in equation \ref{eq:Energy-equation}.

\begin{equation}
S_{latent}={\displaystyle \rho L\frac{dg}{dt}}\label{eq:latent-heat-term}
\end{equation}
\nomenclature[aL]{$L$}{Latent heat}\nomenclature[ag]{$g$}{Volume fraction of solid}\nomenclature[gphis]{$\phi_s$}{Volume fraction of mush}\nomenclature[afs]{$f_s$}{Volume fraction of solid in mush}

with $g$ the volume fraction of solid material. While the heat source
is active, $g$ is evaluated as the equilibrium change of volume fraction
of solid $f_{s}$ 

\begin{equation}
f_{s}=\frac{T_{l}-T}{T_{l}-T_{s}},\, T_{s}<T<T_{l}\label{eq:latent-heat-solid-fraction-fs}
\end{equation}

During the solidification stage, when the heat source is switched
off, the solid fraction $g$ is determined from non-equilibrium solidification
kinetics as a combination of the growth of equiaxed and columnar dendrites.
The envelope fraction (volume fraction of mush) within a control volume
is now given by $\phi_{s}=\phi_{col}+\phi_{equi}$. The solid fraction
within that envelope (the area shaded gray in figure \ref{fig:mushy-zone-schematic})
is again denoted by $f_{s}$. Now, the solid fraction $g$ within
a control volume follows from 

\[
g=\phi_{s}f_{s}
\]

Equation~\ref{eq:latent-heat-solid-fraction-fs} has been implemented
as a linear evolution of the solid fraction between $T_{l}$ and $T_{s}$,
in order to reduce the computational complexity of the model \citep{Browne2004Fixed}
(avoiding the need for iteration), as in this preliminary model the
focus is on successfully coupling both FT and CFD models. A more physically
realistic, non-linear relationship, e.g. the Lever Rule or the Scheil
equation \citep{Duggan2015Modelling}, could be used at the expense
of computational efficiency.

\nomenclature[aTsl]{$T_s$, $T_l$}{Solidus and liquidus temperature}

\subsubsection{Coupling of momentum and heat transport}

Through the inclusion of the momentum sink term, the momentum equation
\ref{eq:Navier-Stokes} is valid for the entire domain including both
liquid and solid regions. In the (semi-)solid regions, a distinction
is made if it is composed of a columnar or equiaxed crystal structure.
Columnar dendrites are stationary as they are attached to the unmolten
solid, and thus are modeled as a porous medium, introducing a momentum
sink following the isotropic Blake-Kozeny model~\citep{Singh2001Modelling}

\[
\vec{F}_{damp}=\frac{\mu K}{\rho}\vec{U}
\]

\[
K=K_{0}\frac{g^{2}}{(1-g)^{3}+\epsilon}
\]
\\
with $\mu K_{0}=\SI{e6}{\newton\second}$ and $\epsilon=10^{-3}$.

Equiaxed grains are able to move with the liquid, and with increasing
equiaxed volume fraction the liquid metal will turn into a slurry.
We model this by increasing the viscosity in the slurry, following
a correlation suggested by \citet{Thomas1965Transport} for spherical
solid particles in a liquid:
\begin{alignat}{1}
\mu_{equi}=\mu\Bigl[ & 1+2.5\phi_{equi}f_{s}+10.05(\phi_{equi}f_{s})^{2}+\nonumber \\
 & \,0.00273\exp(16.6\phi_{equi}f_{s})\Bigl]\label{eq:thomas_slurry_viscosity}
\end{alignat}

This model breaks down when the equiaxed network becomes coherent
at high volume fractions. Thus, we switch to the porous medium model
for equiaxed regions with a high volume fraction above the coherency
threshold.

\subsection{Boundary conditions}

We assume the weld pool to be axisymmetrical and make use of this
by only simulating a wedge of the domain. Circumferential gradients
and velocities are zero on the wedge faces. The boundary conditions
for the other faces are outlined in the following.

\subsubsection{Heat input}

At the top surface, the laser irradiation is modeled by a Gaussian
distributed heat flux. At all surfaces, including the top surface,
there is an outflux of heat due to natural convection and radiation.
This heat outflux is very small compared to the laser irradiation,
but once the latter is switched off, it is this heat outflux which
together with conduction of excess heat into the unmolten material
causes solidification of the weld.

\[
\lambda\frac{\partial T}{\partial n}=\dot{q}_{laser}-\dot{q}_{radiation}-\dot{q}_{convection}
\]

\selectlanguage{ngerman}%
\addtocounter{equation}{-1}\begin{subequations}\vspace{-5mm}

\selectlanguage{english}%
\begin{eqnarray}
\dot{q}_{radiation} & = & \sigma_{b}\epsilon(T^{4}-T_{\infty}^{4})\label{eq:bc_rad}\\
\dot{q}_{convection} & = & h(T-T_{\infty})\label{eq:bc_conv}\\
\dot{q}_{laser} & = & k_{q}\frac{\eta P}{\pi r_{q}^{2}}\exp(-k_{q}\frac{r^{2}}{r_{q}^{2}})\label{eq:bc_laser_heat}
\end{eqnarray}

\selectlanguage{ngerman}%
\end{subequations}

\selectlanguage{english}%
\nomenclature[aq]{$\dot{q}$}{Heat flux}\nomenclature[gsigmab]{$\sigma_b$}{Stefan Boltzman constant}\nomenclature[gepsilon]{$\epsilon$}{Emmisivity}\nomenclature[ah]{$h$}{Convective heat transfer coefficient}\nomenclature[aTinf]{$T_{\infty}$}{Ambient temperature}\nomenclature[akq]{$k_q$}{Gaussian distribution coefficient}\nomenclature[geta]{$\eta$}{Laser absorptivity}\nomenclature[aP]{$P$}{Laser power}\nomenclature[arq]{$r_q$}{$1/e^2$ radius for Gaussian distribution}

The numerical values of the input parameters are given in table~\ref{tab:Heat-input-BC}.

\subsubsection{Momentum}

At the top gas-liquid interface, we introduce a shear stress in the
liquid due to surface tension gradients (i.e. Marangoni forces) along
the interface:

\begin{equation}
\mu\nabla_{n}U_{t}=\frac{d\gamma}{dT}\nabla_{t}T\label{eq:Marangoni-BC}
\end{equation}

The variation of surface tension with temperature is computed using
the thermochemical model of \citet{Sahoo1988Surface}. 

At all other surfaces, we set the velocity to zero. \nomenclature[ggamma]{$\gamma$}{Surface tension}\nomenclature[xsubscriptt]{$t$}{Tangential direction}\nomenclature[xsubscriptn]{$n$}{Normal direction}

\section{Numerical procedure}

Our solver is built on top of the open source finite volume framework
OpenFOAM (version 2.1.x) \citep{Weller1998Tensorial}. Postprocessing
of the results for publication is done with the open source software
matplotlib \citep{Hunter2007Matplotlib}. 

The non-linearity associated with the pressure-velocity-coupling is
handled by the iterative PISO algorithm \citep{Issa1986Solution}.
Once a divergence free velocity field has been computed, the temperature
equation is solved. If a phase change occurs, the temperature equation
will be non-linear and thus solved iteratively. The non-linearity
due to latent heat during melting is dealt with using an implicit
source term linearization technique \citep{Voller1991GENERAL}. Here,
we assume instantaneous progression of the melting front based on
local temperature conditions. While this algorithm is appropriate
for melting conditions, during solidification it fails to predict
undercooling and the metallurgically relevant transition from columnar
to equiaxed grain growth. This deficiency is alleviated by using the
mesoscale front-tracking model for solidification predictions by Browne
and coworkers \citep{Browne2004Fixed}. In this case we displace the
computational markers forming the solidification front based on the
temperature of a previous time step, and then iteratively determine
the current temperature and solid fraction within the mushy zone ($f_{s}$
in equation \ref{eq:latent-heat-solid-fraction-fs}) with under-relaxation.

Since OpenFOAM can perform 3D simulations only, the solution domain
is an axisymmetric wedge of size $r=\SI{10}{\milli\meter}$ and $z=\SI{5}{\milli\meter}$
with an opening angle of \ang{5}. The axisymmetric model presents
a marked improvement over previously published 2D planar results e.g.
by \citet{Duggan2012Combined} or \citet{Koseki2003Numerical}. For
the fluid flow and heat transfer equations, we use a mesh of 200x100
uniform cells. For the volume averaged approach to determine equiaxed
nucleation and growth we use a mesh that is coarse enough to ensure
a reasonable amount of grain refining particles per control volume.
Here, we use a uniform mesh of 160x80 cells, corresponding to an average
of four grain refining particles per grid cell at $\rho_{seeds}=1000^{3/2}\si{\per\milli\meter\cubed}$.
The code is arranged in such a way that the mesh for the fluid flow
and heat transfer calculations can be chosen independently from the
mesh used for the nucleation and growth calculations. The computations
on the two meshes are fully coupled, meaning that data (temperature
from macro- to meso-, and solid fraction from meso- to macroscale)
is exchanged in both directions at every time step via a linear interpolation
framework~\citep{TongMultiScaleMintweld}. This allows for an independent
adaptation of the fluid flow and heat transfer mesh to the mesh requirements
imposed by the macroscopic length scales.

The time step is fixed at \SI{25}{\micro\second}. We use a 2nd order
backward differencing time marching scheme, and a 2nd order TVD scheme
(limitedLinear) for the divergence terms.

\section{Results and discussion}

Following the previous studies by \citet{Villafuerte1990Effect,Villafuerte1995Mechanisms},
\citet{Park2011Effect}, \citet{Duggan2012Combined} and \citet{Koseki2003Numerical,Koseki2002ColumnarToEquiaxed},
we investigate a laser spot weld on an {Fe-15.9wt\%-Cr-14.1wt\%-Ni}
steel alloy. To determine the growth kinetics coefficient $C$, this
ternary alloy is treated as a pseudo-binary alloy, as the chromium
segregates preferentially to the liquid and the partition coefficient
of the nickel is close to 1.0, and thus the chromium is assumed to
be the dominant solute \citep{Elmer1990Singlephase}. 

In contrast to previously published computational results, we include
fluid flow in our analysis, as it is known to have a strong effect
on the heat transfer and shape evolution in a weld pool, both during
the melting stage and the solidification stage. The influence of fluid
flow on the mesoscale grain structure formation during solidification
in welding has not been reported in the literature to date.

\subsection{Weld evolution during the melting stage}

The fluid flow in the pool is driven by surface tension gradients
due to temperature gradients. The surface tension dependency on temperature
is tightly coupled to the concentration of surface active agents (surfactants)
present in any weld pool, such as sulfur and oxygen. In the following
we neglect the presence of oxygen and assume sulfur is the only surfactant
(homogeneously) present in the system.

A sketch of the problem investigated is shown in figure~\ref{fig:Laser-welding-schema}.
The relevant material properties together with their respective literature
sources are listed in table~\ref{tab:Material-properties}. The properties
related to heat inputs and losses are listed in table~\ref{tab:Heat-input-BC}.
For simulations with grain refining particles present, unless otherwise
specified we use a TiN particle density of $(1000/\si{\milli\meter\squared})^{3/2}$,
which is the 3D equivalent of a 2D density of 1000/\si{\milli\meter\squared}
as used by \citet{Koseki2003Numerical}. In their simulations, \citeauthor{Koseki2003Numerical}
assign a homogeneous fixed size of grain refining particles based
on a measured required undercooling of $\Delta{}T_c=\SI{1.8}{\kelvin}$
by \citet{Bramfitt1970Effect} to achieve nucleation of equiaxed grains
on TiN particles. Here, we assume a log-normal distribution of particle
sizes (equation~\ref{eq:Cumulative-density-function}) as measured
by \citet{Park2011Effect} with parameters $\mu_{1/2}=\SI{0.636e-6}{\meter}$
and $\sigma_{d}=0.1$, such that the peak of the distribution is close
to the value determined by \citeauthor{Bramfitt1970Effect}, while
the spread of the distribution lies in-between the fixed size assumed
by \citeauthor{Koseki2003Numerical} and the wider log-normal distribution
measured by \citeauthor{Park2011Effect}. 

\begin{table*}
\protect\caption{Material properties of the Fe-Cr-Ni alloy\label{tab:Material-properties}}

{\small{}}%
\begin{tabular*}{17cm}{@{\extracolsep{\fill}}lrlc}
\toprule 
{\small{}Property} & {\small{}Value} & {\small{}Unit} & {\small{}Ref.}\tabularnewline
\midrule
{\small{}Eutectic temperature $T_{s}$} & {\small{}$1679$} & {\small{}\si{\kelvin}} & {\small{}\citep{Duggan2012Combined}}\tabularnewline
{\small{}Liquidus temperature $T_{l}$} & {\small{}$1710$} & {\small{}\si{\kelvin}} & {\small{}\citep{Duggan2012Combined}}\tabularnewline
{\small{}Specific heat capacity $c_{p}$} & {\small{}$780$} & {\small{}\si{\joule\per\kilo\gram\per\kelvin}} & {\small{}\citep{Duggan2012Combined}}\tabularnewline
{\small{}Density $\rho$} & {\small{}$7250$} & {\small{}\si{\kilo\gram\per\cubic\meter}} & {\small{}\citep{Duggan2012Combined}}\tabularnewline
{\small{}Thermal conductivity $\lambda$} & {\small{}$35$} & {\small{}\si{\watt\per\meter\per\kelvin}} & {\small{}\citep{Duggan2012Combined}}\tabularnewline
{\small{}Latent heat of fusion $L$} & {\small{}$1.854\cdot10^{5}$} & {\small{}\si{\joule\per\kilo\gram}} & {\small{}\citep{Duggan2012Combined}}\tabularnewline
{\small{}Dynamic viscosity $\mu$} & {\small{}$6.1625\cdot10^{-3}$} & {\small{}\si{\pascal\second}} & {\small{}\citep{Brooks2005Measurement}}\tabularnewline
{\small{}Surface tension temperature coefficient $\partial\gamma/\partial T$} & {\small{}$-4.3\cdot10^{-4}$} & {\small{}\si{\newton\per\meter\per\kelvin}} & {\small{}\citep{Sahoo1988Surface}}\tabularnewline
{\small{}Entropy factor} & {\small{}$3.18\cdot10^{-3}$} & {\small{}$-$} & {\small{}\citep{Sahoo1988Surface}}\tabularnewline
{\small{}Entropy of segregation} & {\small{}$-1.66\cdot10^{8}$} & {\small{}\si{\joule\per\kilo\gram}} & {\small{}\citep{Sahoo1988Surface}}\tabularnewline
{\small{}Surface excess at saturation} & {\small{}$1.3\cdot10^{-8}$} & {\small{}\si{\kilo\mol\per\square\meter}} & {\small{}\citep{Sahoo1988Surface}}\tabularnewline
{\small{}Burden and Hunt growth kinetics coefficient $C$} & {\small{}$8.995\cdot10^{-5}$} & {\small{}\si{\meter\per\second\per\square\kelvin}} & {\small{}\citep{Duggan2015Modelling}}\tabularnewline
{\small{}Diffusivity of Cr in the liquid $D_{l}$} & {\small{}$1.8\cdot10^{-9}$} & {\small{}\si{\meter\squared\per\second}} & {\small{}\citep{Duggan2015Modelling}}\tabularnewline
{\small{}Gibbs-Thomson coefficient $\Gamma$} & {\small{}$3.88\cdot10^{-7}$} & {\small{}\si{\meter\kelvin}} & {\small{}\citep{Duggan2015Modelling}}\tabularnewline
{\small{}Liquidus slope $m$} & {\small{}$-2.9$} & {\small{}\si{\kelvin}/wt\%} & {\small{}\citep{Duggan2015Modelling}}\tabularnewline
{\small{}Partition coefficient $\alpha$} & {\small{}$0.86$} & {\small{}-} & {\small{}\citep{Duggan2015Modelling}}\tabularnewline
{\small{}Solid-liquid interfacial tension $\gamma_{sl}$} & {\small{}$0.299$} & {\small{}\si{\joule\per\square\meter}} & {\small{}\citep{Duggan2012Combined}}\tabularnewline
{\small{}Fusion entropy $\Delta S_{v}$} & {\small{}$7.71\cdot10^{5}$} & {\small{}\si{\joule\per\kilo\gram\per\kelvin}} & {\small{}\citep{Duggan2012Combined}}\tabularnewline
\bottomrule
\end{tabular*}{\small \par}

\end{table*}

\begin{table}
\protect\caption{Properties related to the laser and heat losses\label{tab:Heat-input-BC}}
{\small{}}%
\begin{tabular*}{8.5cm}{@{\extracolsep{\fill}}>{\raggedright}p{4cm}rc}
\toprule 
{\small{}Property} & {\small{}Value} & {\small{}Unit}\tabularnewline
\midrule
\midrule 
{\small{}Laser power $P$} & {\small{}$1000$} & {\small{}\si{\watt}}\tabularnewline
{\small{}Distribution coefficient $k_{q}$} & {\small{}$2$} & {\small{}$[-]$}\tabularnewline
{\small{}Beam radius $r_{q}$} & {\small{}$2.5\cdot10^{-3}$} & {\small{}\si{\meter}}\tabularnewline
{\small{}Incidence time} & {\small{}$0.8$} & {\small{}\si{\second}}\tabularnewline
{\small{}Absorptivity $\eta$} & {\small{}$1$} & {\small{}$[-]$}\tabularnewline
{\small{}Emissivity $\epsilon$} & {\small{}$0.5$} & {\small{}$[-]$}\tabularnewline
{\small{}Convective heat transfer coefficient $h$} & {\small{}$8$} & {\small{}\si{\watt\per\meter\squared\per\kelvin}}\tabularnewline
{\small{}Ambient temperature $T_{\infty}$} & {\small{}$300$} & {\small{}\si{\kelvin}}\tabularnewline
\bottomrule
\end{tabular*}{\small \par}

\end{table}

\begin{figure}
\includegraphics{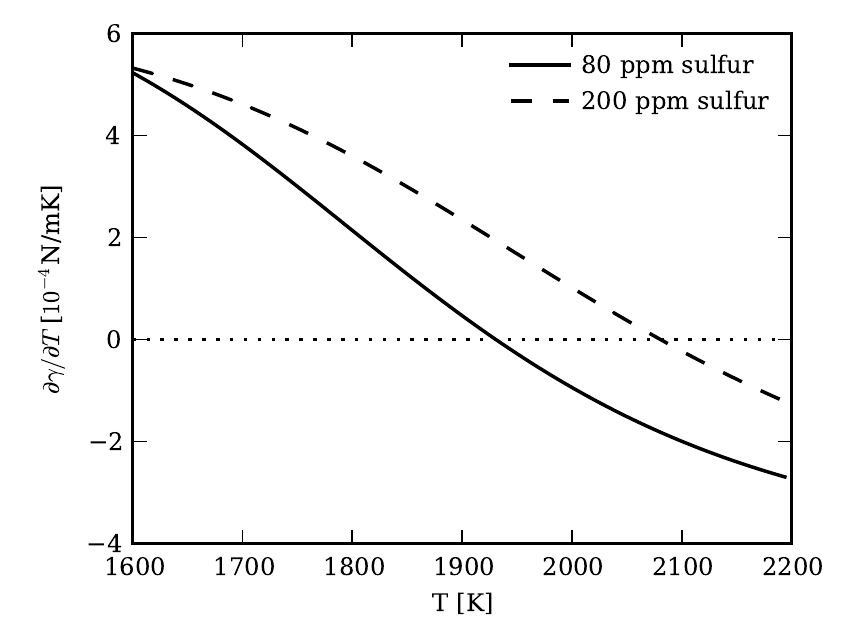}

\protect\caption{Change of surface tension temperature coefficient with temperature
and impurity concentration\label{fig:surface-tension-temperature}}

\end{figure}

\begin{sidewaysfigure*}
\includegraphics[bb=0bp 25bp 230bp 185bp,clip]{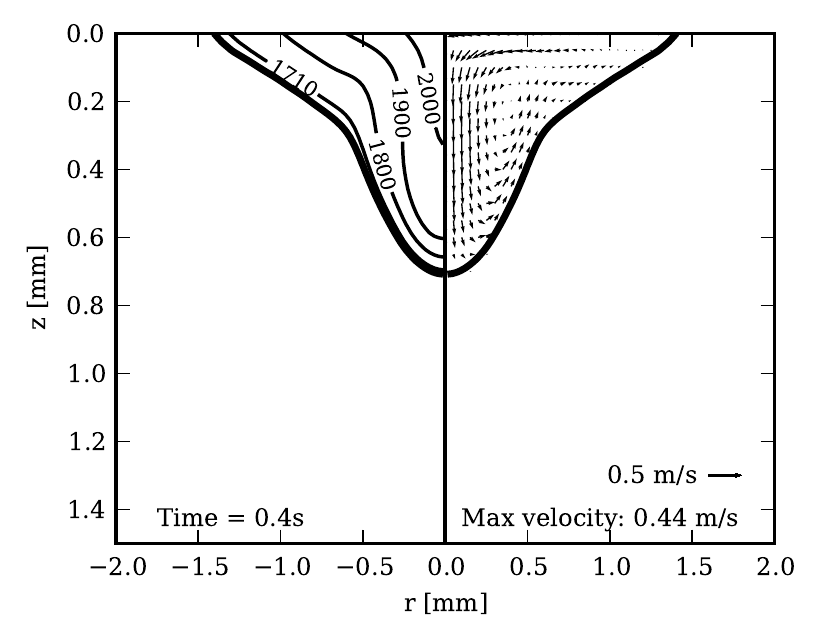}\includegraphics[bb=18bp 25bp 230bp 185bp,clip]{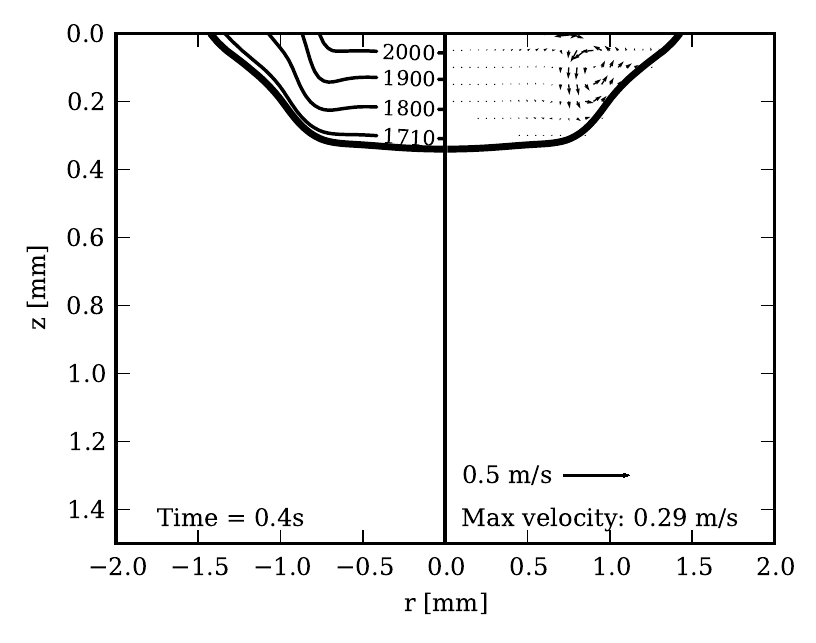}\includegraphics[bb=18bp 25bp 230bp 185bp,clip]{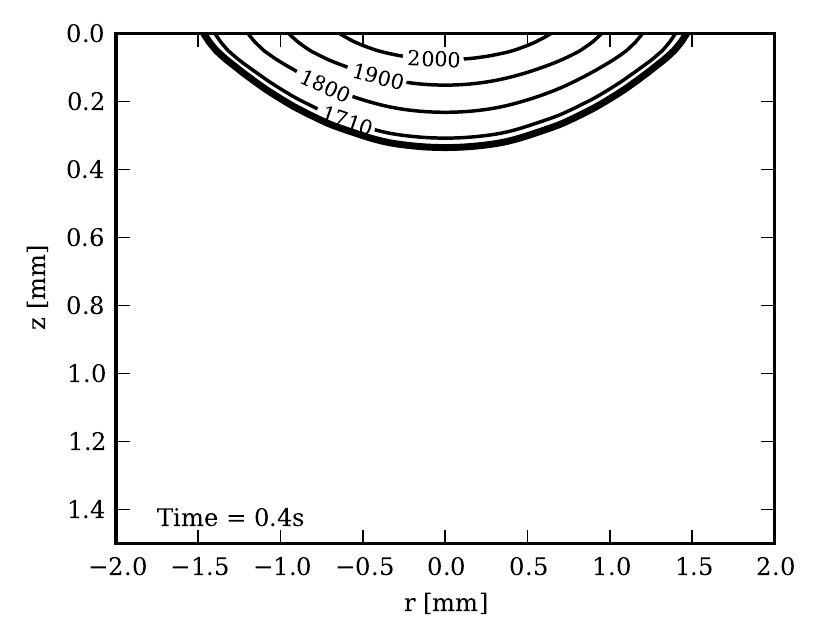}

\includegraphics[bb=0bp 25bp 230bp 185bp,clip]{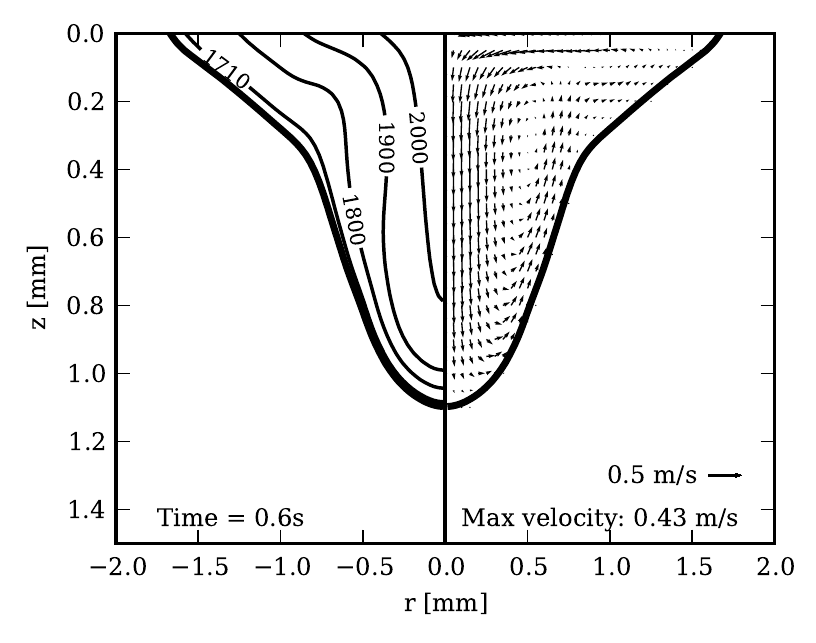}\includegraphics[bb=18bp 25bp 230bp 185bp,clip]{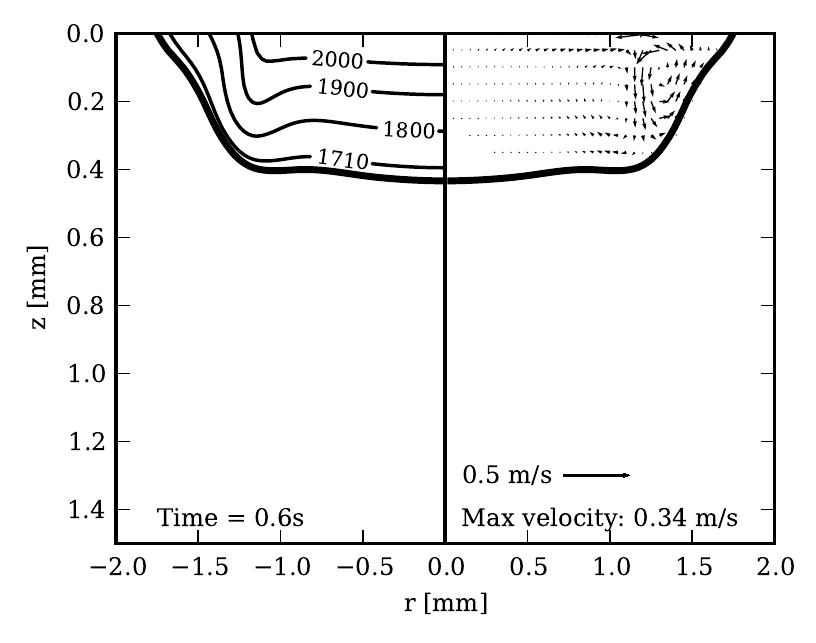}\includegraphics[bb=18bp 25bp 230bp 185bp,clip]{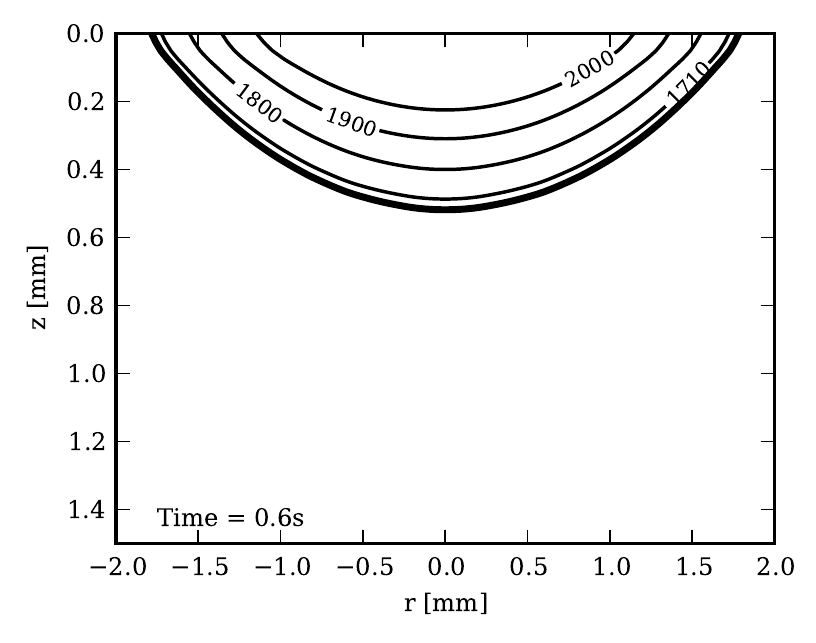}

\includegraphics[bb=0bp 2bp 230bp 185bp,clip]{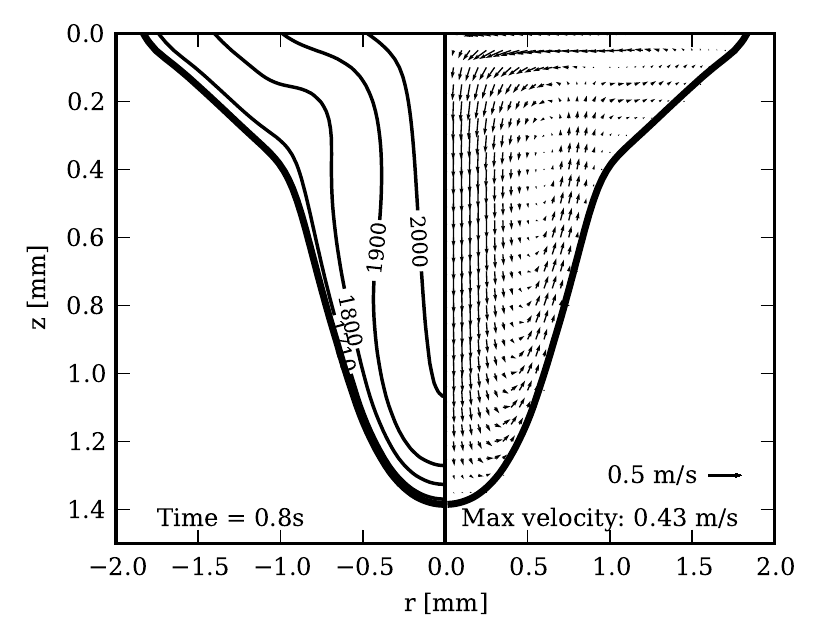}\includegraphics[bb=18bp 2bp 230bp 185bp,clip]{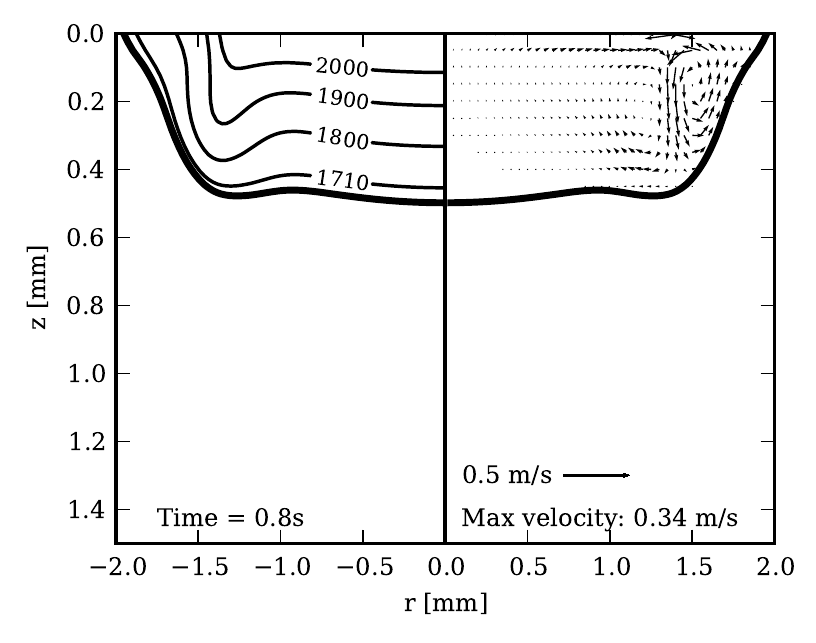}\includegraphics[bb=18bp 2bp 230bp 185bp,clip]{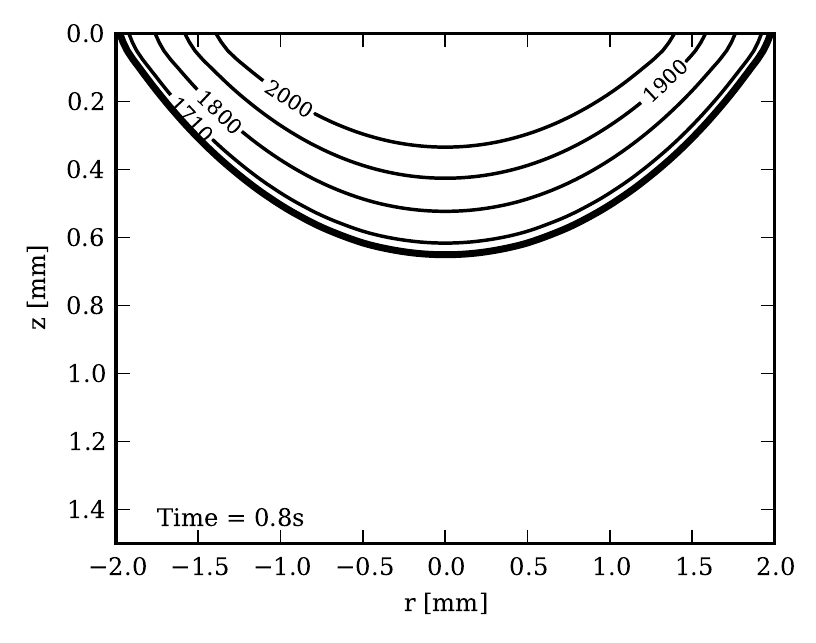}

\protect\caption{Development of the weld pool with isotherms in Kelvin and flow velocity
vectors during the heating stage at three time instances with 200~ppm
sulfur (left column), 80~ppm sulfur (middle column) and pure diffusion.
The final image shows the state at t=0.8s when the laser is shut off
and solidification begins.\label{fig:initial_shape_and_flow}}
\end{sidewaysfigure*}

The high power laser irradiation will lead to melting of the base
metal in fractions of a second and thus the formation of a weld pool.
For an alloy with a sulfur concentration of 200~ppm the surface tension
gradient ${\displaystyle \partial\gamma/\partial T}$ is positive
for temperatures up to nearly \SI{2100}{\kelvin} (figure~\ref{fig:surface-tension-temperature}).
Since fluids flow towards regions of higher surface tension, the fluid
flow will dominantly be directed towards the center of the pool, where
the highest temperatures lead to the highest surface tension. Such
a flow pattern leads to a weld pool with a width over depth ratio
smaller than 1, characteristic for conduction mode welds (a weld with
negligible vaporization of the melt \citep{Lienert2011Laser}) with
a high content of surface active species (figure~\ref{fig:initial_shape_and_flow},
left column). At lower sulfur concentrations, the sign change of ${\displaystyle \partial\gamma/\partial T}$
is shifted towards lower temperatures (figure~\ref{fig:surface-tension-temperature}).
For a concentration of 80~ppm, ${\displaystyle \partial\gamma/\partial T}$
will not remain positive over the entire weld pool surface, and thus
the maximum surface tension will not be located at the center of the
weld pool but shifted radially outwards, close to the critical \SI{1900}{\kelvin}
isotherm. The liquid metal in the weld pool will flow towards this
point, leading to two recirculation zones in the weld pool, resulting
in decreased penetration with a characteristic bulge of the pool boundary
under the stagnation point and a width to depth ratio larger than
1 (figure~\ref{fig:initial_shape_and_flow}, middle column). The
computed maximum velocities in the weld pool are in the order of  0.3-{\SI{0.4}{\meter\per\second}},
which is in good agreement with a maximum velocity of $[(\partial\gamma/\partial T)q/(\mu c_{p})]^{0.5}=\SI{0.3}{\meter\per\second}$
obtained by a scaling analysis \protect\cite{Chakraborty2002Scaling},
and simulation results of \SI{0.4}{\meter\per\second} for comparable
conduction mode laser welds \protect\cite{PitschenederDebRoy}.

Here it is illustrative to stress that neglecting fluid flow entirely,
as has been done in previous studies \citep{Villafuerte1990Effect,Duggan2012Combined,Duggan2012Integrated,Koseki2003Numerical,Zhan2008Dendritic},
leads to a very different, hemispherical weld pool shape, irrespective
of the sulfur concentration (figure~\ref{fig:initial_shape_and_flow},
right column). This highlights the need of a proper weld pool flow
model to obtain an accurate initial condition for the solidification
computations.

\subsection{Weld evolution during the solidification stage}

Whereas the final shape of the weld is mostly (but not entirely) determined
during the heating phase, its grain morphology is fully determined
during the cooling phase. Here we study the influence of accounting
for solidification kinetics using our front tracking model versus
the use of an equilibrium enthalpy method, and the influence of accounting
for fluid flow, on the shape and morphology evolution during the cooling
stage.

\begin{figure}
\hfill{}\includegraphics[bb=0bp 0bp 231bp 185bp]{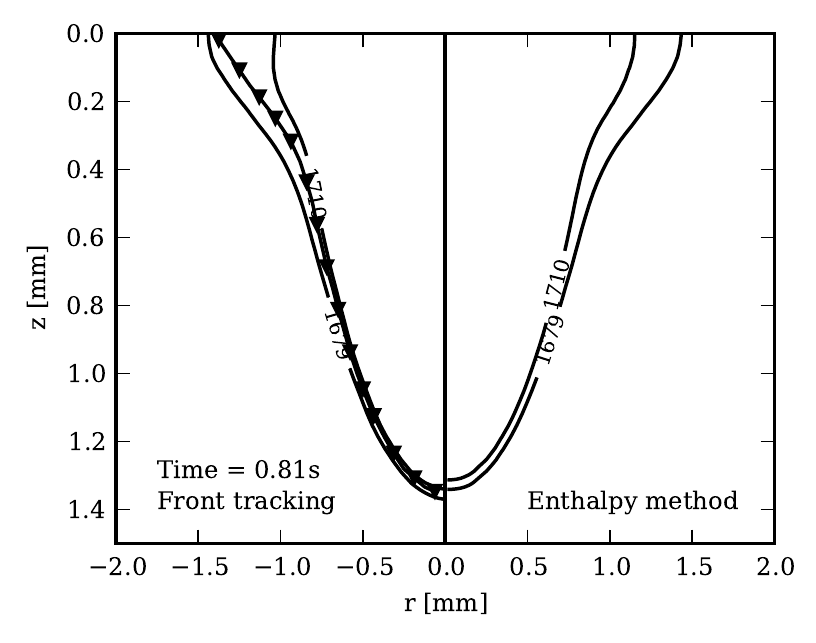}\hfill{}

\hfill{}\includegraphics[bb=0bp 0bp 231bp 185bp,width=8cm]{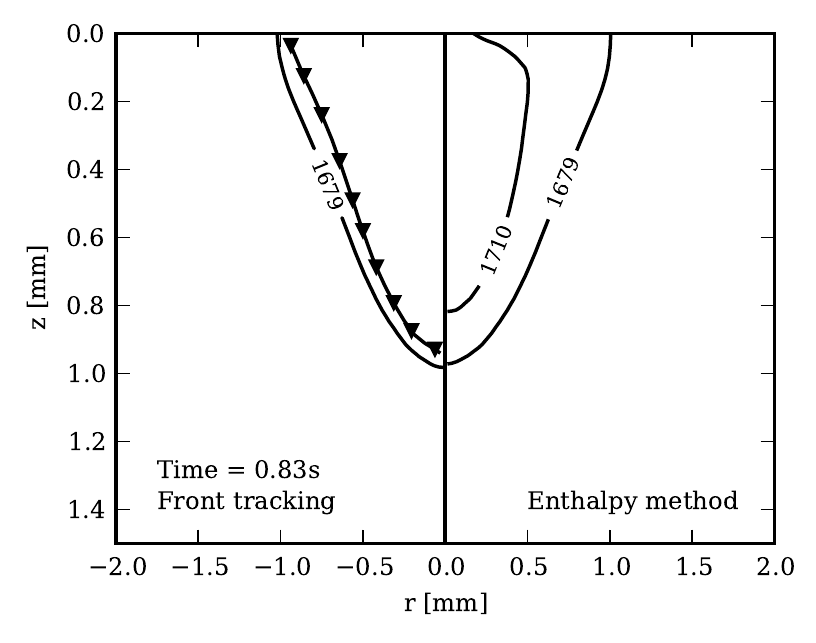}\hfill{}

\protect\caption{Temperature evolution at two time instances after the heat source
has been switched off at t=0.80s, for 200~ppm sulfur concentration.
The location of the columnar dendritic solidification front is depicted
by a line with triangle symbols. \label{fig:Temperature-evolution}}
\end{figure}

\begin{figure}
\hfill{}\includegraphics[bb=0bp 0bp 231bp 185bp]{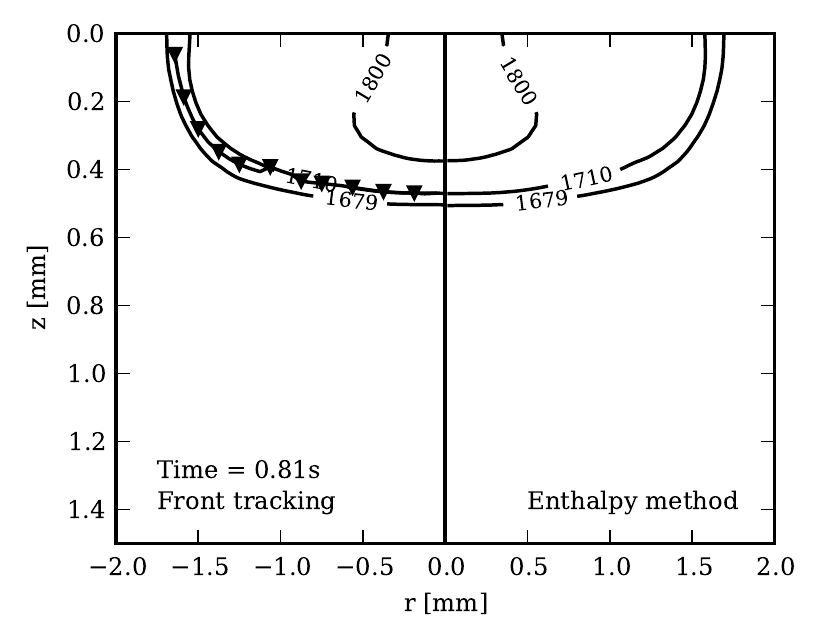}\hfill{}

\hfill{}\includegraphics[bb=0bp 0bp 231bp 185bp,width=8cm]{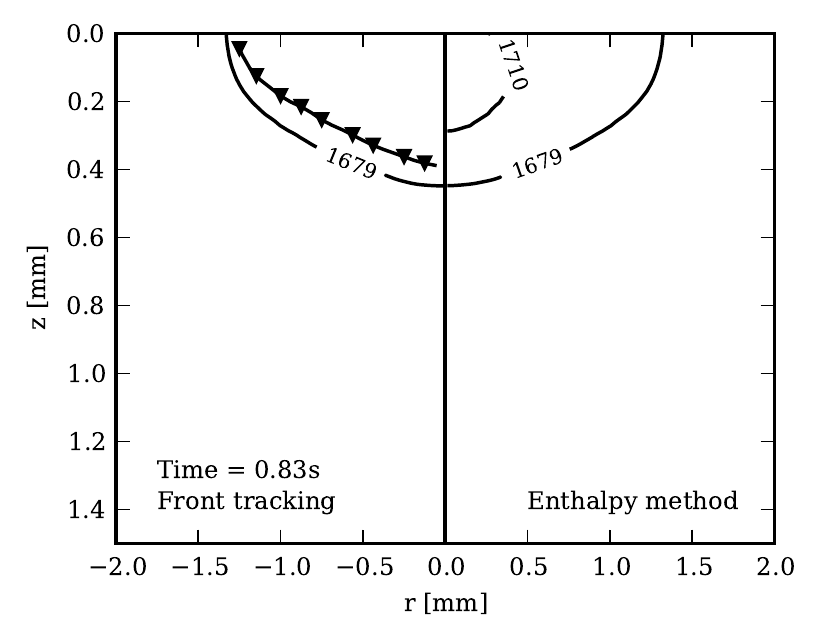}\hfill{}

\protect\caption{Temperature evolution at two time instances after the heat source
has been switched off, 80~ppm sulfur\label{fig:LS-Temperature-evolution}}
\end{figure}

For a 200~ppm sulfur weld, the evolution of the melt pool and temperatures
therein, at two time instances after deactivation of the heat source,
is shown in figure~\ref{fig:Temperature-evolution}, computed both
with the commonly used enthalpy method which does not take into account
growth kinetics, and with our combined front-tracking equiaxed growth
model. With the latter the temperatures within the weld pool show
some difference when compared to the computed temperatures using an
enthalpy method. Shown are the $T_{s}=\SI{1679}{\kelvin}$ solidus
temperature isotherm and the $T_{l}=\SI{1710}{\kelvin}$ liquidus
isotherm. Notable differences are found in the location of the liquidus
isotherm and these differences increase during further solidification.
In the enthalpy method the solidification front is coincident with
the liquidus isotherm, whereas in the front-tracking method the solidification
front progression is significantly delayed resulting in a large undercooled
region and a much thinner mushy zone. The same conclusion can be drawn
for the completely differently shaped weld in a low (80~ppm) sulfur
alloy (see figure~\ref{fig:LS-Temperature-evolution}). The transient
position of the solidus isotherm with the two models, on the other
hand, is virtually the same for both the high and low sulfur case. 

Thus, simulations aimed solely at predicting the weld pool shape can
be carried out with a basic enthalpy method without the need for a
sophisticated solidification model. If the post-solidification grain
structure is of interest however, a sophisticated solidification model,
such as the one presented here, is necessary to compute accurate thermal
gradients and solidification front progression, as enthalpy methods
have no microstructure information.

\begin{figure}
\hfill{}\includegraphics{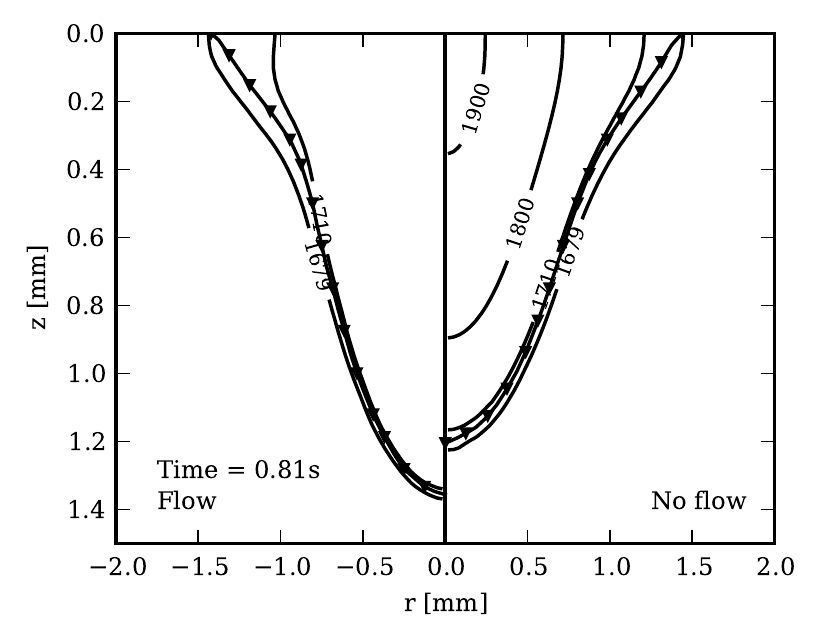}\hfill{}

\hfill{}\includegraphics{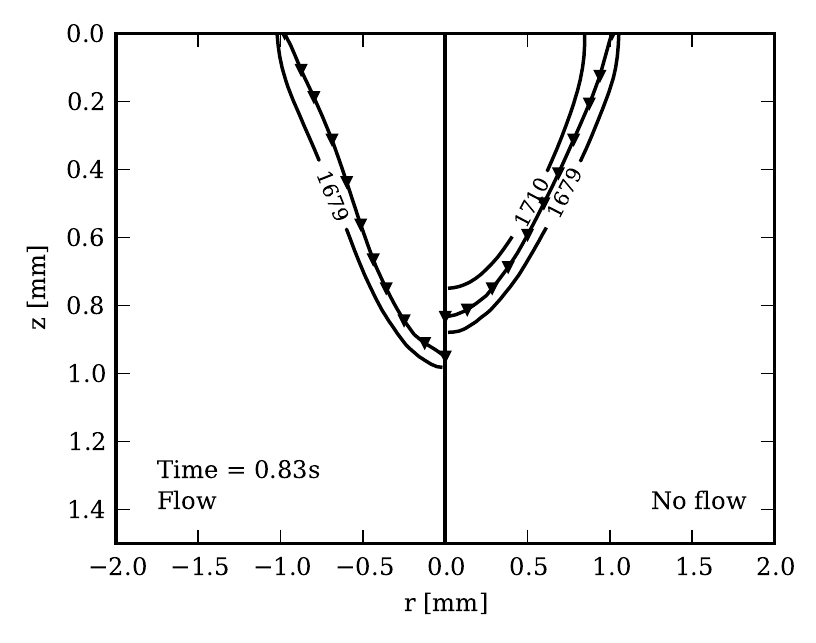}\hfill{}

\protect\caption{Temperature evolution at two time instances during solidification
with and without the consideration of fluid flow, 200~ppm sulfur.
The line with triangle symbols depicts the location of the columnar
dendritic solidification front.\label{fig:Temperature-evolution-flow-noflow}}
\end{figure}

\begin{figure}
\hfill{}\includegraphics{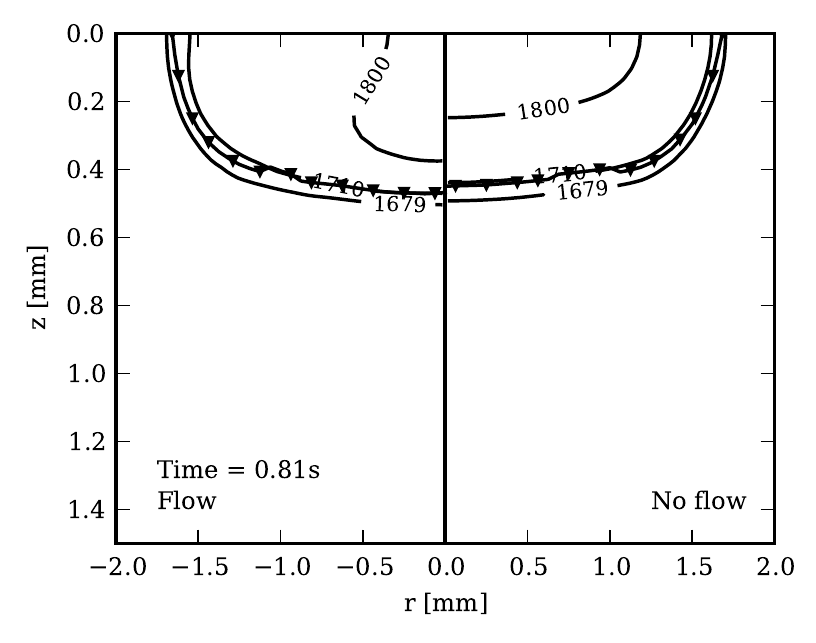}\hfill{}

\hfill{}\includegraphics{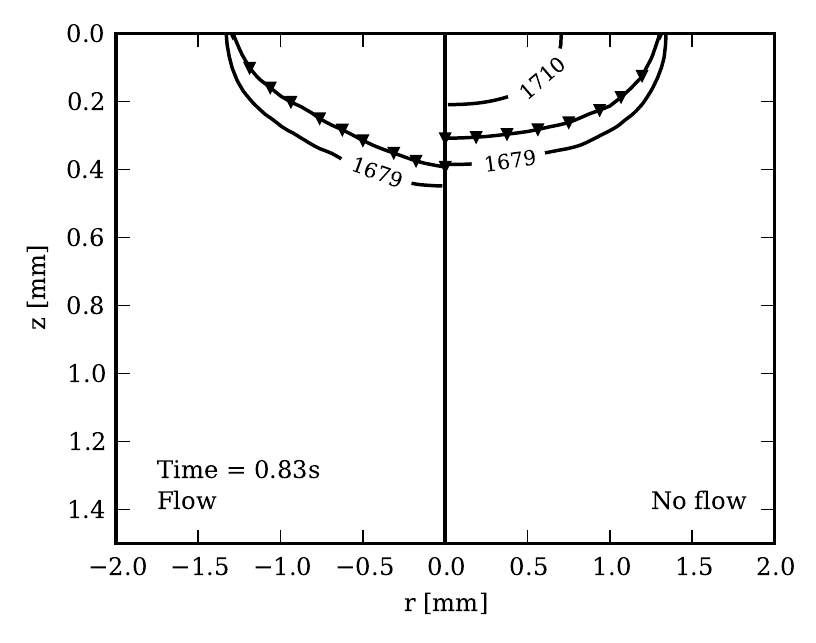}\hfill{}

\protect\caption{Temperature evolution at two time instances during solidification
with and without the consideration of fluid flow, 80~ppm sulfur.
The line with triangle symbols depicts the location of the columnar
dendritic solidification front. \label{fig:LS-Temperature-evolution-flow-noflow}}
\end{figure}

The question of the significance of fluid flow during the cooling
phase on the evolution of the solidification process is addressed
in figures~\ref{fig:Temperature-evolution-flow-noflow} and \ref{fig:LS-Temperature-evolution-flow-noflow},
for high and low sulfur concentrations respectively. For both cases,
starting from the same initial condition at the end of the melting
phase, the solidification phase is subsequently computed with and
without fluid flow. The inclusion of fluid flow in the solidification
phase leads to better mixing and heat transfer, and thus lower and
more uniform temperatures in the core of the weld pool. This leads
to significant differences in the weld pool shape \citep{Ehlen2003Simulation,Saldi2013Effect},
and more importantly in the post-solidification grain morphology,
as will be shown in the next section. 

For the high sulfur case, the inclusion of fluid flow, which in this
case is directed inward along the weld top surface and downward along
the weld axis, leads to continued heat transport to the bottom of
the weld after the heat source has been switched off, and thus to
continued local melting. This causes a deeper and less wide weld as
compared to the situation in which fluid flow is ignored.

For the low sulfur case, the downward fluid flow and convective heat
transport is weaker and further away from the axis. Yet, the inclusion
of fluid flow leads to continued heat transport to the bottom of the
weld after the heat source has been switched off, and a deeper weld
as compared to the situation in which fluid flow is ignored.

\subsection{Influence of grain refining particles on grain morphology evolution
during solidification}

Up to this point, we have investigated solidification within the weld
pool in the absence of grain refining particles, leading to columnar
dendritic growth only. When taking into account the promotion of equiaxed
nucleation using grain refiners, at some point the columnar front
progression is blocked and the solidification structure transitions
into equiaxed grains. This is addressed in the current subsection.

\subsubsection{Sensitivity to numerical and model parameters}

First, we assess the sensitivity of the main simulation outcome of
interest, \emph{viz.} the location of the columnar-equiaxed transition
line, as represented by the total volume of equiaxed solid in the
weld, on \emph{(i)} time step and mesh size and \emph{(ii)} the precise
value of the threshold for coherence of equiaxed grains.

Ad \emph{(i)}: A simulation with a doubled fixed time step of \SI{50}{\micro\second}
resulted in a slight increase of 4.6\% of the volume of equiaxed solid
(and a corresponding decrease in the volume of columnar dendritic
solid). The total volumes of equiaxed solid in the weld, as computed
with a fixed time step of \SI{25}{\micro\second} on three different
meshes, \emph{viz.} the standard mesh, and two meshes that are coarsened
and refined by a linear factor 1.25 compared to the standard mesh
(see \citet{Roache1994Perspective} for a justification of the refinement
ratio), show a monotonic convergence, allowing for the computation
of an asymptotic value by Richardson extrapolation. Compared to the
asymptotic value, the coarse mesh result is 6.1\% smaller, the base
mesh result 2.0\% smaller and the fine mesh result 0.48\% smaller.

\begin{figure}
\hfill{}\includegraphics[bb=0bp 0bp 239bp 185bp]{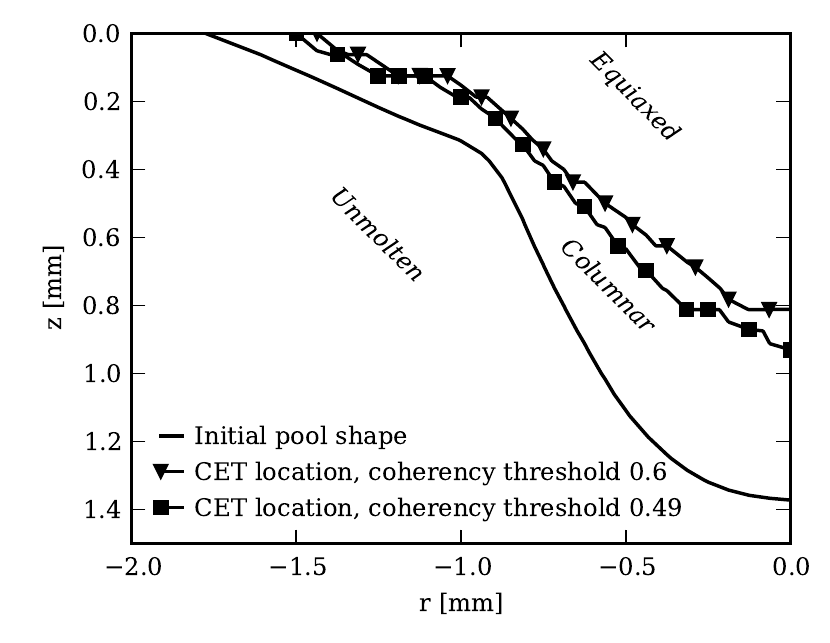}\hfill{}

\protect\caption{Sensitivity on the coherency threshold of the computed transition
from columnar to equiaxed solidification structure in a weld pool,
200~ppm sulfur\label{fig:CET-fcoh}}
\end{figure}

Ad \emph{(ii)}: In a previous section, we mentioned that the precise
value of the threshold for coherence of the equiaxed grains is not
agreed upon in the literature (ranging from $0.2$ \citep{Biscuola2008Mechanical}
to $1$ \citep{Mirihanage2010Macroscopic}). To study the sensitivity
of our results to this threshold value, we simulated solidification
of a weld pool using a coherency threshold of $0.49$ and $0.6$,
leading to a 15\% difference in the volume of equiaxed structure in
the weld (figure~\ref{fig:CET-fcoh}). The same conclusion has been
drawn by \citet{Mirihanage2009Combined}, who found the occurrence
of the columnar-to-equiaxed transition (CET) in a casting to be rather
insensitive to the exact value of the coherency threshold. All further
simulations in the current study were performed with the most commonly
used blocking threshold of $0.49$ as proposed by \citet{Hunt1984Steady}.

\subsubsection{Influence of grain transport}

\begin{figure}
\hfill{}\includegraphics[bb=0bp 0bp 239bp 185bp]{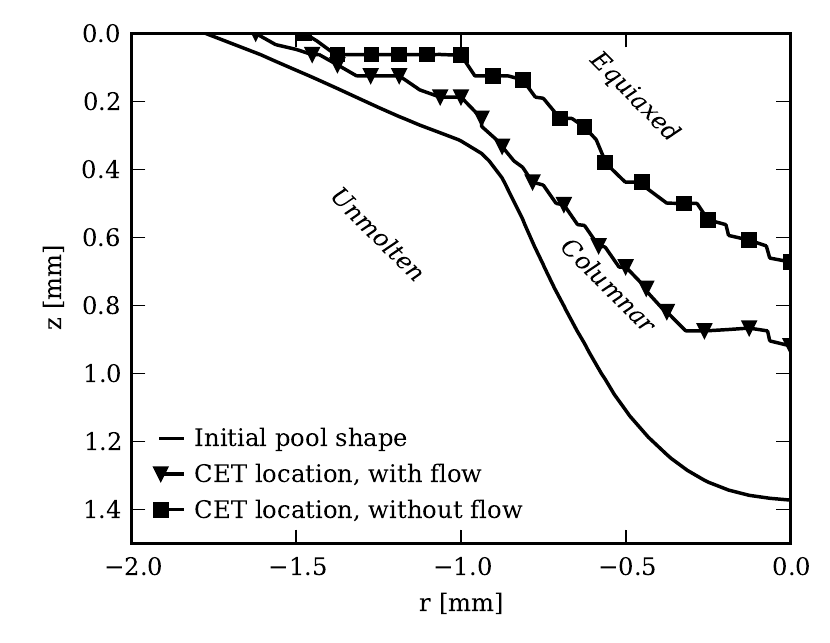}\hfill{}

\protect\caption{Equiaxed transition in a weld pool with and without fluid flow, 200~ppm
sulfur\label{fig:CET-flow}}
\end{figure}

\begin{figure}
\hfill{}\includegraphics[bb=0bp 0bp 239bp 185bp]{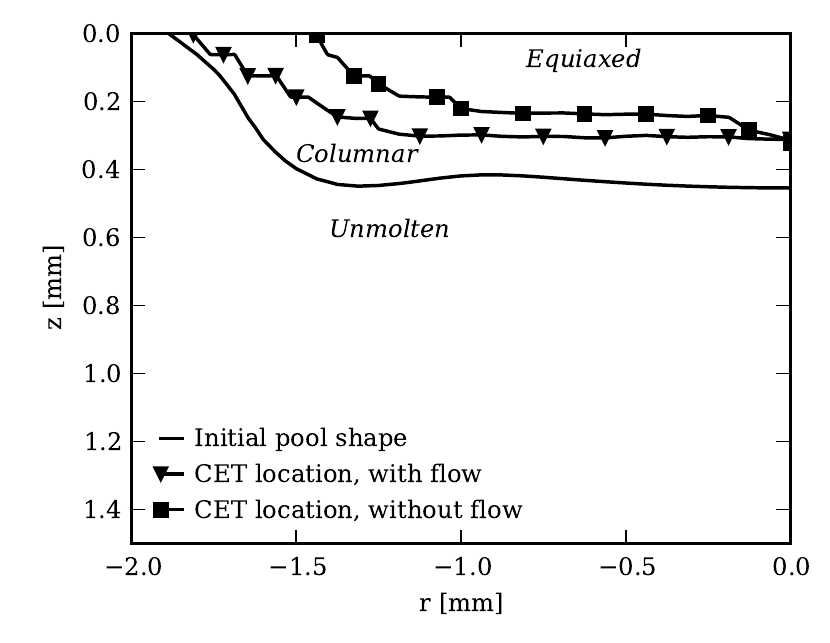}\hfill{}

\protect\caption{Equiaxed transition in a weld pool with and without fluid flow, 80~ppm
sulfur\label{fig:LS-CET-flow}}
\end{figure}

The presence of fluid flow in the weld pool during solidification
leads to a significantly earlier transition to equiaxed growth, as
can be seen in figures~\ref{fig:CET-flow} and \ref{fig:LS-CET-flow}
for high and low sulfur cases respectively. Conversely, the location
of the columnar-to-equiaxed transition was found to be rather insensitive
to the inclusion or neglect of the transport of equiaxed grains with
the flow, as described by equations \ref{eq:grain_transport_equation1}
and \ref{eq:grain_transport_equation2}.

Hence, the earlier transition to equiaxed growth when including fluid
flow is mainly due to the lower thermal gradients and lower temperatures
in the melt (see figure~\ref{fig:Temperature-evolution-flow-noflow}),
which result in a larger liquid region with sufficient undercooling
for the equiaxed nucleation and growth to take place. This conclusion
agrees with Hunt's criterion \citep{Hunt1984Steady}, stating that
the fraction of equiaxed grains is inversely proportional to the magnitude
of the thermal gradients cubed in the melt.

\subsubsection{Influence of grain refining particle number density}

\begin{figure}
\hfill{}\includegraphics[bb=0bp 0bp 239bp 185bp]{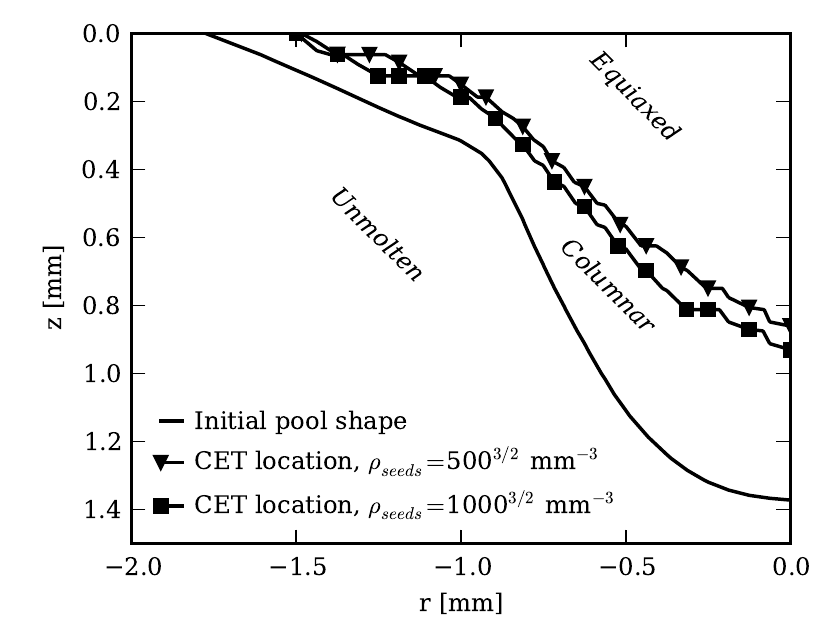}\hfill{}

\protect\caption{Equiaxed transition in a weld pool with two different densities of
grain refining particles, 200~ppm sulfur\label{fig:CET-particle-density}}
\end{figure}

\begin{figure}
\hfill{}\includegraphics[bb=0bp 0bp 239bp 185bp]{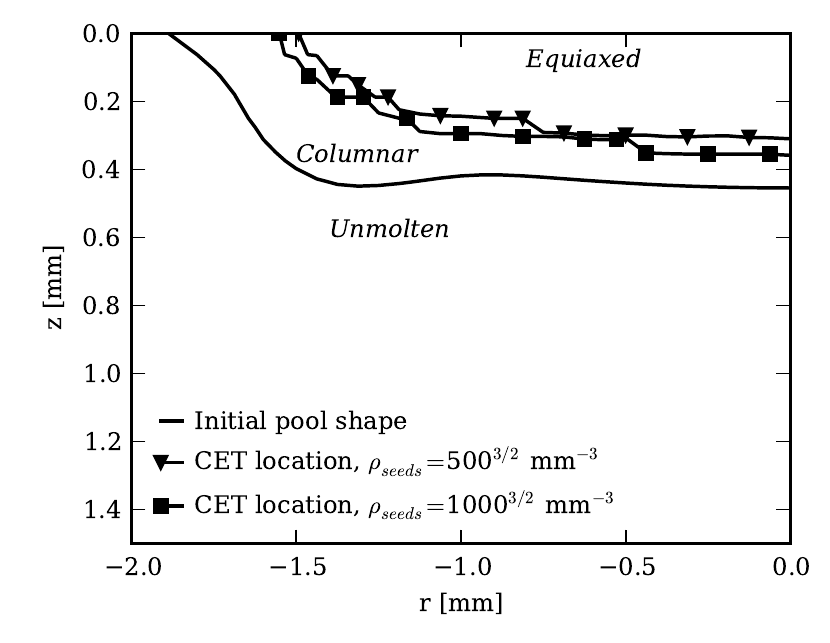}\hfill{}

\protect\caption{Equiaxed transition in a weld pool with two different densities of
grain refining particles, 80~ppm sulfur\label{fig:LS-CET-particle-density}}
\end{figure}

\begin{figure}
\hfill{}\includegraphics[bb=0bp 0bp 239bp 185bp]{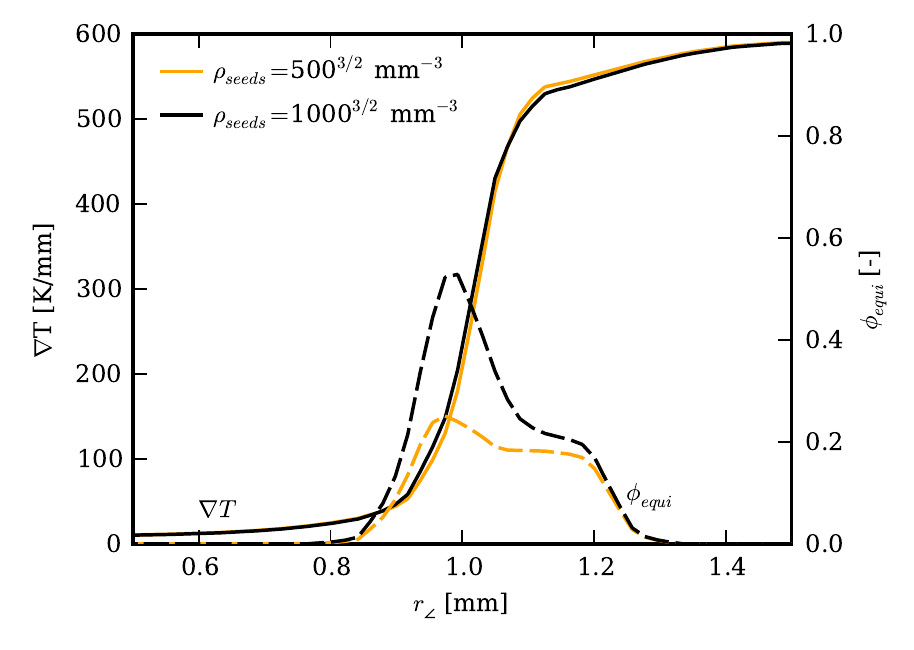}\hfill{}

\protect\caption{Thermal gradient and equiaxed envelope volume fraction along a diagonal
line $z=r$ away from the weld pool center, 200~ppm\label{fig:HS-gradT}}
\end{figure}

\begin{figure}
\hfill{}\includegraphics[bb=0bp 0bp 239bp 185bp]{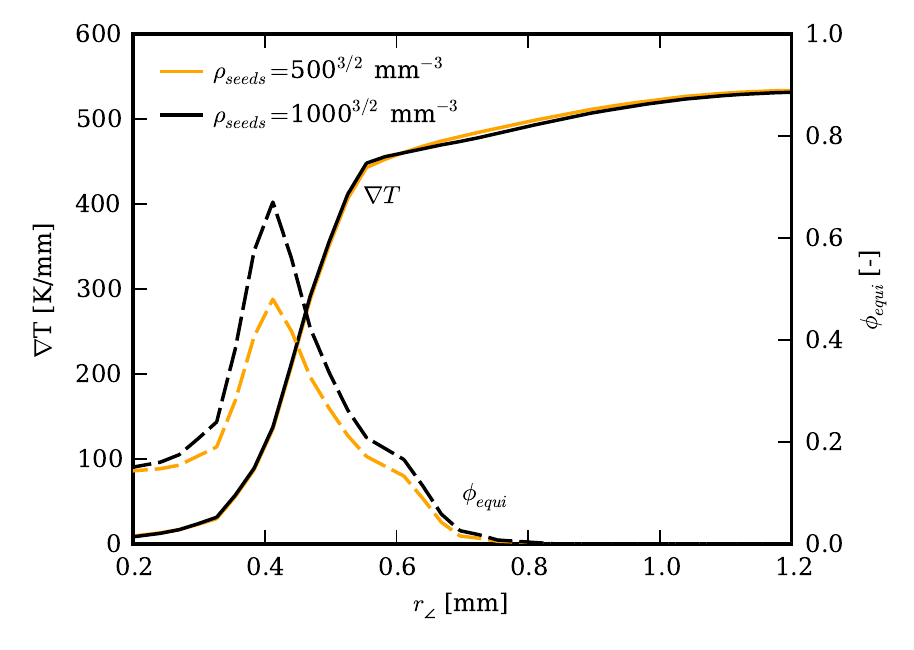}\hfill{}

\protect\caption{Thermal gradient and equiaxed envelope volume fraction along a diagonal
line $z=r$ away from the weld pool center, 80~ppm\label{fig:LS-gradT}}
\end{figure}

The influence of the number density of grain refining particles on
the transition from columnar to equiaxed solidification is illustrated
in figures~\ref{fig:CET-particle-density} and \ref{fig:LS-CET-particle-density},
which show that the transition to equiaxed solidification can be triggered
with a grain refiner density of $500^{3/2}$ particles per \si{\milli\meter\cubed},
and that the position of the transition is only slightly affected
by an increase of the grain refiner number density to $1000^{3/2}$
particles per \si{\milli\meter\cubed}. This observation matches the
findings by \citet{Koseki2003Numerical} that grain refiner density
mainly influences the size of the resulting equiaxed grains, and not
the point at which they reach coherency. Simulating the effect of
a grain refiner density lower than $500^{3/2}$ particles per \si{\milli\meter\cubed}
is not feasible using our current volume averaged approach, as this
would require very large computational cells in relation to the weld
pool area to allow for a sufficient number of grain refiner particles
per computational cell. In figures~\ref{fig:HS-gradT} and \ref{fig:LS-gradT},
we show the thermal gradient and equiaxed envelope volume along a
diagonal line $r_{\angle}=\sqrt{r^{2}+z^{2}}$, $z=r$ at a time instance
where the equiaxed grains reach coherency. We can identify a low,
fairly uniform temperature gradient within the weld pool which rises
sharply around the liquid-solid interface and is insensitive to the
number of grain refining particles present. As a result, favorable
conditions for the growth of equiaxed grains are met at the same location
for both particle densities, though coherency is achieved slightly
later in time for the lower number density. 

\section{Conclusion}

In addition to the macroscopic shape of a weld, the mesoscopic grain
morphology is crucial in determining the resultant properties. These
two are determined by the interaction between macroscale fluid flow
and heat transfer, and mesoscale grain solidification kinetics, during
the heating and cooling phase of a welding process. Whereas both macroscale
and mesoscale phenomena have previously been studied separately through
computational models, an integrated model for both scales is needed
to predict weld properties on both relevant scales.

Such an integrated model has been presented in the current work. The
model is conceptually simple and computationally inexpensive, while
granting insight into the post-solidification shape and grain morphology
of a weld.

The model has been applied to study the melting and solidification
of a laser spot weld on two Fe-Cr-Ni steel alloys, \emph{viz}. with
a low and high sulfur content. Due to their very different dependency
of surface tension on temperature, the weld pools of these two alloys
represent the two extremes in surface tension driven flow commonly
observed during welding. Although demonstrated here for laser spot
welding, the presented results are relevant for arc welding as well,
as both types of welding are equivalent once the heat source is turned
off.

The predicted solidification evolution has been compared to that obtained
with the commonly used equilibrium enthalpy method, in which solidification
takes place quasi-statically and its kinetics is ignored. Apart from
the obvious fact that an equilibrium method cannot predict the important
transitions in grain morphology, a marked difference between the two
methods was observed in the thermal gradients within the weld pool
during solidification. This in turn has a strong impact on the developing
grain structure. However, if only the weld pool shape, and not its
grain morphology, is of interest, the equilibrium enthalpy method
is found to be sufficient to obtain accurate results for the partial
penetration cases examined here.

Neglecting fluid flow during the cooling phase, as done in previously
reported studies, was found to have a significant influence on the
predicted weld pool shape and grain morphology. Thus any accurate
simulation of weld pools should include fluid flow.

The use of grain refining particles was found to be an effective means
to favorably alter the grain morphology by initiating a transition
from columnar to equiaxed growth. This transition was found to be
not very sensitive to the most uncertain parameter in the model, \emph{viz}.
the coherency threshold value, nor to the density of grain refining
particles above a certain critical value.

\section*{Acknowledgments}

We would like to thank the European Commission for funding the MINTWELD
project (reference 229108) via the FP7-NMP program.

\bibliographystyle{elsarticle-num-names}
\bibliography{ucdcouple}

\end{document}